\documentclass[aps,prl,twocolumn,showpacs,preprintnumbers,amsmath,amssymb]{revtex4}

\usepackage{graphicx}
\usepackage{dcolumn}
\usepackage{bm}
\usepackage[usenames]{color}    

\newcommand{\be}{\begin{equation}}
\newcommand{\ee}{\end{equation}}
\newcommand{\bea}{\vspace{0.25cm}\begin{eqnarray}}
\newcommand{\eea}{\end{eqnarray}}

\def\PRL{{Phys. Rev. Lett.} }
\def\PRA{{Phys. Rev.} A }

\begin{document}
\title{Measurement of sub-shot-noise spatial correlations without
    subtraction of background }
\author{Giorgio Brida\textsuperscript{1}, Lucia Caspani\textsuperscript{2}, Alessandra Gatti\textsuperscript{2}, Marco
Genovese\textsuperscript{1}, Alice Meda\textsuperscript{1}, Ivano
Ruo Berchera\textsuperscript{1}} \affiliation{$^1$INRIM, strada
delle Cacce 91, 10135 Torino, Italy.\\
$^2$ INFM-CNR-CNISM, Universit\'a dell'Insubria, Via Valleggio 11,
22100 Como, Italy.}
\begin{abstract}
In this paper we present the measurement of sub-shot-noise spatial
correlations without any subtraction of background, a result paving
the way to realize sub-shot-noise imaging of weak objects.
\end{abstract}
 \maketitle
Quantum properties of the optical field represent a resource of the
utmost relevance for the development of quantum technologies,
allowing unprecedented results in disciplines ranging from quantum
information and metrology \cite{zei-mg} to quantum imaging (for
recent reviews see \cite{kol-gatti2008}). For applications in this
latter field,  a fundamental tool to be realized is the spatial
correlation, at the level of quantum fluctuations, of two optical
beams. An example is the high-sensitivity detection of weak objects
proposed in \cite{brambilla2008}, which exploits the intrinsic
multi-mode  quantum correlations  of parametric down conversion
(PDC) emission, a result that may have disruptive practical
applications in biomedical imaging or whenever there is the need of
illuminating an object with a low flux of photons \footnote{For
example, low illumination levels are needed in presence of pigments
or other photo-sensible molecules, see e.g. \cite{lp}. The  sub-shot
noise imaging requires a photon flux roughly $\sigma $ times smaller
than an equivalent classical one,
 in order to achieve the same signal-to-noise ratio
\cite{brambilla2008}.}. This technique is based on the spatial
correlation in the quantum noise of two conjugated branches of PDC
emission \cite{gatti1999,brambilla2001,brambilla2004}. The image of
a weak absorbing object in one branch, eventually previously hidden
in the noise, can be restored by subtracting the strongly correlated
spatial noise pattern measured in the other branch. In order to
achieve a sensitivity superior than that available with classical
techniques, one should be able to reach a sub-shot-noise (SSN)
regime in spatial correlations even in the presence of the
unavoidable background noise (electronic noise, scattered
light)\cite{brambilla2008}. Nevertheless, up to now, such a result
was not yet achieved. Indeed, previous demonstrations of the quantum
nature \cite{dit, dit2} of the spatial correlation of PDC  beams
were realized only for low photon numbers. In that regime the
background noise was dominant, so that a proof-of-principle
demonstration of  SSN correlation was possible only after correcting
the results for the background noise (i.e. by subtracting the
variance of the spatial pattern of the background, measured in the
absence of PDC). Clearly, such a regime cannot be used for concrete
imaging schemes, as the image distribution would remain hidden in
the background noise. Similarly, a single-mode sub-shot-noise
intensity correlation \cite{8,9,10,11,12,masha} cannot be used to
retrieve high-sensitivity information on the spatial distribution of
an object, since the  quantum character of the correlation vanishes
when one detects small portions of the beams instead of the whole
beams. Thus,  high sensitivity imaging requires spatial quantum
correlations (i.e. SSN intensity correlations between several
portions of two twin beams) in a regime where the photon flux is
high enough to make the background noise negligible. Purpose of the
present letter is to present this achievement: a clear observation
of SSN spatial correlation, without any correction for the
background noise, is presented and discussed, opening the concrete
possibility of realizing imaging with a sensitivity beyond the
standard quantum limit.
\par
The strategy we shall follow to reach our goal is that indicated by
\cite{brambilla2008}: the major limitation of the experiment
\cite{dit} was the presence in each beam of a large excess noise
with respect to coherent light, which made  SSN
correlation fragile with respect to unavoidable experimental
imperfections (unbalances, errors in the determination of the
symmetry center of the signal-idler distributions) and rapidly
deteriorated the correlation for increasing gain. Such an excess
noise can be lowered by working in a different regime, i.e. a pump
pulse of duration much longer than the typical coherence time of PDC
beams, which is on the order of ps for a few mm crystal. As
predicted in \cite{brambilla2008}, a nanosecond pump pulse should
enable the observation of SSN correlations at a much
higher photon number.
\begin{figure}[tbp]
\begin{center}
\includegraphics[ width=9.5 cm, height=6cm, bb=0 0 1200 800]{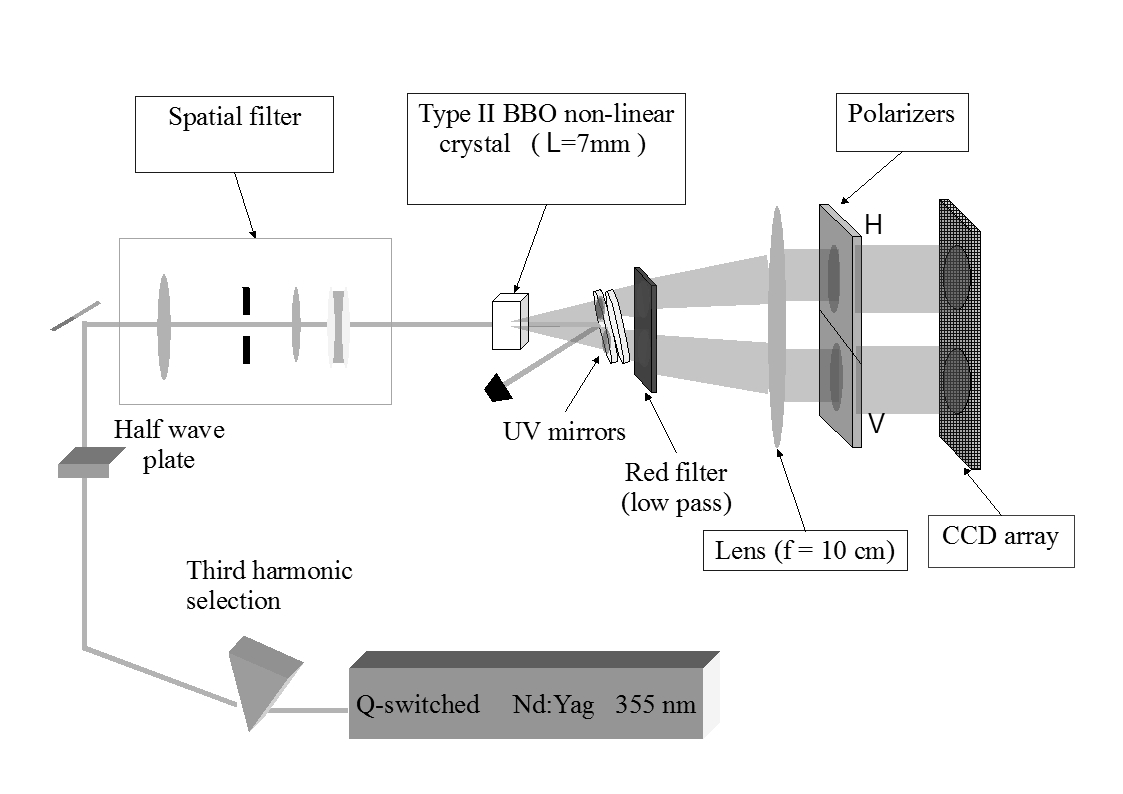}
\caption{Experimental setup. A triplicated Nd-Yag laser beam, after
spatial filtering, produces type II PDC in a BBO crystal, which is
then measured (after eliminating UV beam) by an high sensitivity CCD
camera.}\label{fig1}
\end{center}
\end{figure}
In our setup (Fig.\ref{fig1}) a type II BBO non-linear
crystal ($l=7$ mm) is pumped by the third harmonic (355 nm) of a
Q-switched Nd:Yag laser (with pulses of 5 ns, 10 Hz repetition rate
and a maximum energy of about 200 mJ) after spatial filtering (a
lens with a focal length of 50 cm and a diamond pin-hole, 250 $\mu$m
of diameter), for eliminating the non-gaussian components, and
collimation (by a system of lenses to a diameter of $w_{p}=1.25$
mm). After  eliminating  the pump (by a couple of UV mirrors,
transparent to the visible, T=98\% at 710 nm, and by a low
frequency-pass filter, T=95\% at 710 nm) the far-field of the down
converted beams (signal and idler) is imaged onto a CCD camera, by
means of a lens ($f=10$ cm) in a $f-f$ optical configuration,
ensuring that we image the Fourier transform of the crystal exit
face. Therefore each transverse mode with wavevector $\mathbf{q}$
of the emission is associated to a point $\mathbf{x}=(\lambda
f/2\pi)\mathbf{q}$ of the image. We use a $1340X400$ CCD array,
Princeton Pixis400BR (pixel size of 20 $\mu$m), with high quantum
efficiency (80\%) and low noise in the read out process. The CCD
exposure time is set to 90 ms, thus each image catched in this time
window corresponds to the PDC emission generated by a single shot of
the laser. After having identified the degenerate emissions (two
separated rings in our phase matching configuration), at wavelength
$\lambda_s=\lambda_i=710 nm$, by inserting an interference filter,
we focus the analysis on two correlated areas of side 2,72 mm within
a 10 nm bandwidth around it. In each of the two signal-idler
regions, the unwanted contribution of frequency components of the
orthogonal polarizations other than the degenerate one are
eliminated by a polarizer T=97\%). Notice that differently with the
setup of \cite{dit} the signal and idler beams follow basically the
same optical path, so that unbalances are reduced.
\par
In the ideal case of a plane wave pump ($\mathbf{q}_{pump}=0$), the
momentum conservation in the microscopic PDC process requires that
each pair of signal and idler photons is emitted in symmetric
directions $\mathbf{q}_{s}=-\mathbf{q}_{i}$. As a consequence,
perfect intensity spatial correlation between any symmetric portions
$\mathbf{x}_{s}=-\mathbf{x}_{i}$ of the far-field zone of the signal
and idler beams were predicted \cite{gatti1999,brambilla2001,
brambilla2004}. For a finite pump of waist $w_p$,  having a
distribution of transverse momenta, there is an uncertainty in the
relative directions of propagation of twin photons represented by
the angular bandwidth of the pump, so that the photon numbers
collected from symmetrical portions of the far-field zone are
perfectly quantum correlated only when the detection areas are
broader than a coherence area (roughly the transverse size of the
mode in the far field), whose size $\Delta x_{coh}^2$ depends on the
pump angular bandwidth $\lambda/w_{p}$ \cite{brambilla2001,
brambilla2004}, and on the parametric gain
\cite{dit2,brambilla2008,nos2-nos1}. Thus, we expect a quantum
correlation in the number of photons $\widehat{N}_{i}$ and
$\widehat{N}_{s}$ detected at any couple of symmetric positions with
respect to the center of symmetry, providing our detection areas are
larger than the typical coherence area.
\par
The quantity that characterizes the amount of correlation is the
noise reduction factor $\sigma$, defined as the variance of the
photon number difference $\widehat{N}_{i}-\widehat{N}_{s}$
normalized to the shot noise. In the ideal case of detection areas
much larger than the coherence area we have the usual "twin beams"
result: \cite{gatti1999, josa}
\begin{equation}\label{sigma}
\sigma\equiv\frac{\left\langle\delta
^{2}(\widehat{N}_{i}-\widehat{N}_{s})\right\rangle} {\left\langle
\widehat{N}_{i}+\widehat{N}_{s}\right\rangle}=1-\eta,
\end{equation}
where $\eta$ represents the total transmittance of the optical
channel, including the quantum efficiency of the detectors. For
classical light $\sigma$ is above 1 and reaches the lowest limit in
the case of coherent light ($\sigma_{coh}=1$) which is referred to
as Shot Noise Level (SNL). Whereas for twin beams $\sigma$ is always
smaller than unity (see Eq.\ref{sigma}). The other relevant quantity
is  the Fano factor $F_{s(i)}\equiv\langle\delta
^{2}\widehat{N}_{s(i)}\rangle/\langle \widehat{N}_{s(i)}\rangle$,
which represents the fluctuation of the single beam normalized to
the shot noise. For PDC , since each beam taken alone has a
thermal-like statistics, the Fano factor has the multi-thermal form:
$F_{s(i)}=1+E_{n}$, where $E_{n}=\langle\widehat{N}_{s(i)}\rangle/M$
is called "excess noise" and $M$ is the degeneracy factor,
proportional to the number of spatial and temporal modes collected
in the measurement.
\par
We estimate the expectation values as averages over a large spatial
ensemble (large regions of a single image). Specifically, let us
define $N(\mathbf{x})$ the number of photons, registered by the CCD
pixel in the position $\mathbf{x}$ of a generic region $R$ of the
CCD. $\delta N(\mathbf{x})= N(\mathbf{x})-\langle
N(\mathbf{x})\rangle$ is the fluctuation around the mean value,
estimated as $\langle N(\mathbf{x})\rangle=(1/n)\sum_{\mathbf{x}}
N(\mathbf{x})$, with $n$ the number of pixels in $R$. We estimate
the $\sigma$ parameter by fixing a region  $R_{s}$ in the signal
portion (136X136 pixels), and searching the corresponding region
$R_{i}$ of the idler field that minimizes the experimental noise
reduction factor:
\begin{equation}\label{sigma exp}
 \sigma_{exp}(\mathbf{\xi})=\frac{\left\langle\delta^{2}\left[N_{s}(\mathbf{x})-N_{i}(-\mathbf{x}+\mathbf{\xi})\right]\right\rangle}
 {\left\langle N_{s}(\mathbf{x})+N_{i}(-\mathbf{x}+\mathbf{\xi})\right\rangle}
\end{equation}
where $\mathbf{\xi}$ is the displacement vector of $R_{i}$ in the
pixel space with respect to the optimal position. In other words,
the numerator represents the residual spatial noise of the image in
$R_{s}$  after having subtracted pixel-by-pixel the correlated noise
pattern in  $R_{i}$.
\begin{figure}[tbp]
\begin{center}
\includegraphics[ width=9 cm, height=7 cm, bb=0 0 800 500]{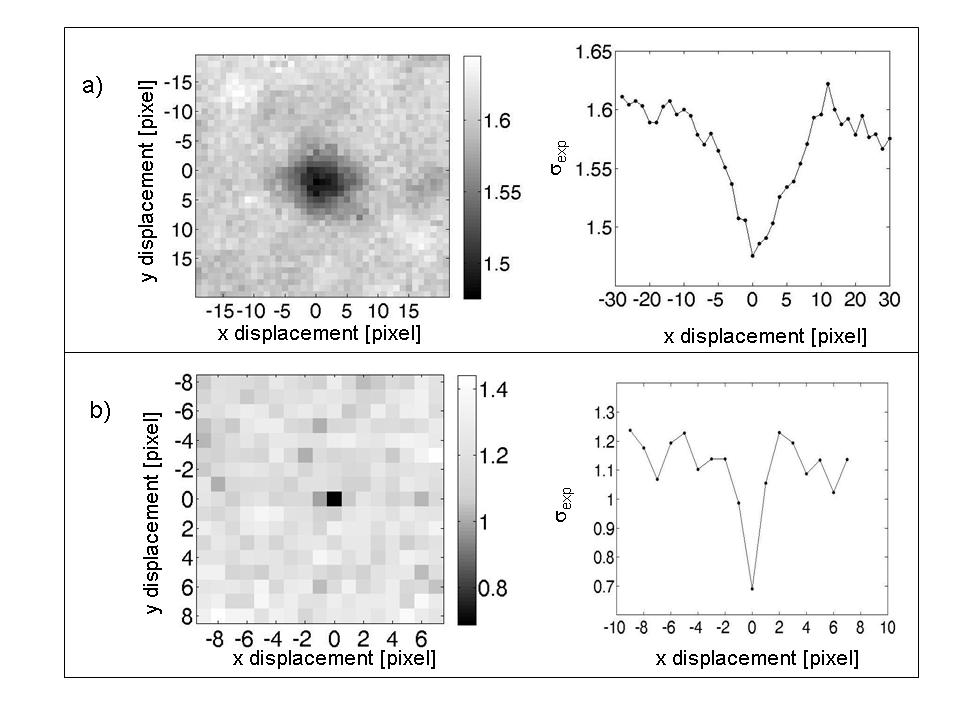}
\caption{ Example of noise reduction factor $\sigma_{exp}$,
evaluated for a specific image,  as a function of the 2D position of
the region $R_{i}$ (left panels).  The right panels are sections of
the minimum. (a) $\sigma$ is measured using the CCD pixels (side 20
$\mu$m). Its minimum value stays just 0.15 below the Fano factor
($\sigma$ evaluated far from the minimum) and is above 1. (b)
$\sigma$ is measured using larger  pixels (side 160 $\mu$m). In this
case, the correlation is higher (the gap between the Fano factor and
the minimum is about 0.6) and,  especially, the minimum is below 1,
i.e.  the correlation is sub-shot-noise }\label{fig2}
\end{center}
\end{figure}
$\sigma_{exp}(\mathbf{\xi})$ is plotted in Fig.\ref{fig2}. Far from
the minimum, we are in the situation in which the two regions are
completely uncorrelated, and  $\sigma_{exp}$ approaches the Fano
factor of the single region (we suppose for simplicity $F_{s}\approx
F_{i}$). Furthermore, Fig.\ref{fig2} (a) shows that the size of the
coherence area is larger than the pixel and the Full Width Half
Maximum of the deep gives us $\Delta x_{coh}$ in terms of pixels.
For this image,  we estimate $\Delta x_{coh}\approx 8$ pixels
(160$\mu m$). Since the SSN can be achieved only for detection areas
containing one or more coherence areas, we performed a binning of
the pixels, i.e. we group the physical pixels in blocks of dimension
$8X8$, called superpixels. The number of photons collected in a
superpixel is the sum of the photons collected by each original
pixel. Our CCD allows to perform the binning operation at the
hardware level, also lowering strongly the signal to electronic
noise ratio. Fig.\ref{fig2} (b) shows $\sigma_{exp}$ in the case of
binning. The deep $\sigma_{exp}(0)$, representing the estimate of
$\sigma$, is noticeably under the SNL while the FWHM of the
coherence area is clearly on the order of the superpixel dimension.
We note that, although the operation of binning reduces the spatial
statistical ensemble, nevertheless the chosen region $R_{s}$ is
large enough to contain about 300 superpixels. This number provides
an estimate of the spatial resolution, because it gives the number
of details of an image that one would be able to distinguish in an
imaging experiment. It is clear \cite{brambilla2008} that the
resolution in this high-sensitivity scheme is limited by the size of
the coherence area of PDC, which in principle can be reduced by
acting on the pump transverse size (a systematic experimental study
of this dependence having been presented in \cite{nos2-nos1}).
\begin{figure}[tbp]
\begin{center}
\includegraphics[ width=8 cm, height=5.5 cm, bb=0 0 1200 800]{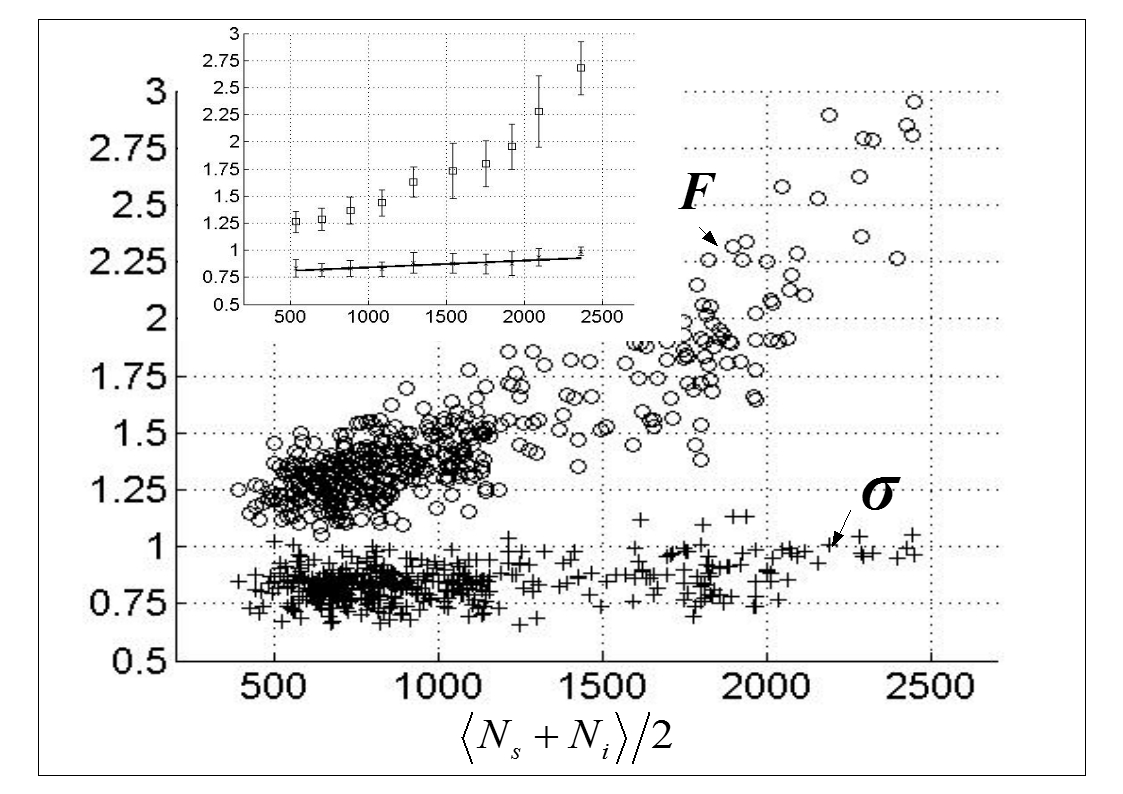}
\caption{ The noise reduction factor $ \sigma_{exp} $, without any
background correction,  is plotted versus the mean photon number per
pixel. Each point (crosses)  in the main window corresponds to the
analysis of a single image. The corresponding Fano factor, $F$, is
reported by the circles. Several images reach a noise reduction
under 0.7. The insert on the top-left angle presents the same data
averaged over tens of images. The equation for the fit is $y=(5.9
\pm 1.5) 10^{-5} x + (0.78 \pm 0.02)$. }\label{fig3}
\end{center}
\end{figure}
\par
Fig.\ref{fig3} presents the minimum values of $\sigma_{exp}$ plotted
as a function of the photon number, where for each data point the
center of symmetry of the two images (the position of the minimum of
$\sigma_{exp}$) is determined with sub-pixel resolution, using the
algorithm described in \cite{caspani2008}. The corresponding Fano
factor $F=(F_{s}+ F_{i})/2$ is also reported. Each point of the main
graph represents the result evaluated for a single image, generated
by single laser shot, while in the small insert the data are
averaged over several images. The comparison between $F$ and
$\sigma$ gives a measure of the reduction of the noise level that
can be observed in a differential measurements, where the
intensities pattern of correlated regions are properly subtracted.
Although no a posteriori correction for experimental noise and
imperfections has been applied, all the fitting line is clearly SSN
in the range of photon number considered, with a minimum of 0.82 for
a photon number per pixel of 600 (the smallest value for a single
shot being 0.66). Also the dispersion of the data is reduced, being
almost all the point under the SNL. This figures are further reduced
to $\sigma=0.7$ when a 12x12 binning is used (the smallest value for
a single shot being 0.5). This two facts represent a fundamental new
feature with respect to \cite{dit}, where a lot of images were above
SNL and a heavy noise correction had been applied on the
experimental values of $\sigma$. Our data demonstrate that in our
setup the spatial fluctuations of two selected regions,
comprehensive of the unavoidable electronic noise and diffuse light,
are really correlated at the quantum level, not only in principle,
and therefore it is possible to exploit this correlation for real
and new application to imaging. For example for an object
transmittance $\alpha<15\%$, following the theoretical prediction of
\cite{brambilla2008} we can estimate that our setup would provide a
SNR more than 10\% better than classical light  with 8x8 binning (20
\% with a 12x12 binning).
\par
Let us consider the sources of noise that can spoil the spatial
correlation. First of all the electronic noise in the read-out and
digitalization process is $\Delta=4$ photoelectron/pixel and
$\Delta=9$ photoelectron/superpixel in the case of binning. Then,
the quantum efficiency of the pixels fluctuates spatially with a
standard deviation $\delta\eta=3\%$. Anyway, this is a very minor
effect in our photon number regime. On the other side, the
fluorescence of the crystal, mirrors and filters, absorbing the high
energy pump pulse, together with the residual light of the pump
itself, contribute to a background of diffuse light collected by the
CCD in each shot. This represents the main source of noise in our
present configuration. For the sake of comparison with previous
experiments, we can correct for this background noise, acquiring an
image where the PDC emission has been suppressed, simply rotating
the pump polarization by 90 degrees. Let us define $S^{2}_{s(i)}$
the variance of the spatial fluctuations of this image in the
regions $R_{s(i)}$, and $M_{s(i)}$ the mean photon numbers. We have
typical values of $S^{2}_{s(i)}\sim200$ and $M_{s(i)}\sim 120$. The
formula for the corrected noise reduction factor is
\begin{equation}\label{sigma exp}
 \sigma_{cor}(0)=\frac{\left\langle\delta^{2}\left[N_{s}(\mathbf{x})-N_{i}(-\mathbf{x})\right]\right\rangle-S^{2}_{s}-S^{2}_{i}}
 {\left\langle N_{s}(\mathbf{x})+N_{i}(-\mathbf{x})\right\rangle-M_{s}-M_{i}}
\end{equation}
\begin{figure}[tbp]
\begin{center}
\includegraphics[width=9cm, height=6 cm, bb=0 0 1200 800]{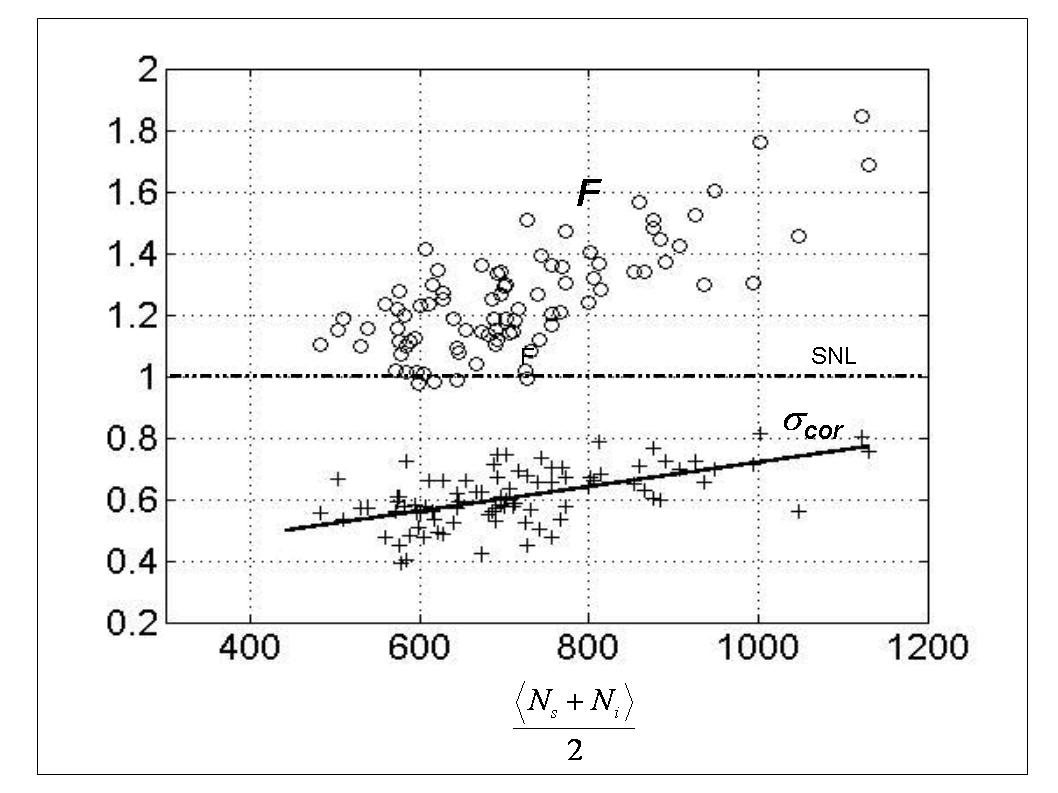}
\caption{Noise reduction factor $ \sigma_{cor}$ and Fano factor $F$,
corrected for the background noise. Here we present a subset of the
same data showed in Fig. \ref{fig4}, once the electronic and the
diffuse light noise has been subtracted. The equation for the fit is
$y=(3.7 \pm 1.1) 10^{-4} x + (0.38 \pm 0.08)$} \label{fig4}
\end{center}
\end{figure}
The same correction can be applied to the Fano factors.
Fig.\ref{fig4} reports the results applied to a subset (for
graphical reasons) of the data of Fig.\ref{fig3}. Some image reaches
values of noise reduction of 0.4 and the average is under 0.6 in the
lower photon number condition. The Fano factors are reduced as well.
These data can be compared with the theoretical threshold for the
noise reduction given by Eq.(\ref{sigma}). Since we estimate in our
setup a total transmission-detection efficiency of 67\%, it leads to
$\sigma=0.33$. This means that we still have  a net margin of
improvement of 0.27. We attribute the gap between the theoretical
prediction and the experimental achievement to i) the fact that in
our setup the superpixel is just  on the order of the coherence
area, while the optimal condition for the observation of SSN would
require a detection area sensibly larger than the coherence area,
and ii) a residual error in the determination of the symmetry center
of the signal-idler distributions, because the bigger is the
superpixel, the  larger is the uncertainty in the evaluation of the
symmetry center, even when this is retrieved by using a sub-pixel
algorithm, as described in \cite{caspani2008}.
\par
In conclusion, in this paper we presented our results showing that
we reached the condition of sub-shot-noise  in spatial correlation
without any background subtraction, a fundamental achievement for
realizing high sensitivity quantum imaging protocols. Indeed, we
obtained a substantial reduction of the noise under SNL in the
spatial distribution of a large portion of PDC emission by
subtracting the correlated one.
\acknowledgments{We acknowledge precious advices and discussions
with E. Brambilla, L.Lugiato, O. Jedrkiewicz. This work was
supported by Compagnia di San Paolo, by PRIN 2007FYETBY, by Regione
Piemonte (E14) and by the EC FET  under the GA HIDEAS
FP7-ICT-221906. We would like to acknowledge that whilst this letter
was under referring process a paper presenting similar results,
albeit in completely different intensity and detection regime and a
configuration not devoted to imaging applications, appeared in this
journal \cite{prl}. }

%
%
%
%

\end{document}